\documentclass[12pt]{iopart}
\usepackage{iopams}

\usepackage{epsfig,psfrag}

\begin{document}
 
\title{High-energetic Cosmic Antiprotons from Kaluza-Klein Dark Matter}

\author{Torsten Bringmann}
\address{Department of Physics, AlbaNova University Center, Stockholm
       University, SE - 106 91 Stockholm, Sweden}
\eads{\mailto{troms@physto.se}}
\date{June 15, 2005}

\begin{abstract}
The lightest Kaluza-Klein particle (LKP) in models with universal extra dimensions is an interesting dark matter candidate that has recently received great attention. Here, we investigate the antiproton flux from LKP annihilations in the galactic halo.
In our analysis we include different halo density profiles and allow for part of the dark matter to be concentrated in 'clumps' rather than being distributed homogeneously. After re-analyzing the observational bounds on the allowed amount of clumpiness, we find that LKP annihilations may well give a significant contribution to the antiproton flux at energies higher than about 10 GeV, while for energies above around 500 GeV the conventional background is expected to dominate again. The shortly upcoming PAMELA satellite will already be able to measure part of this high-energy window, while planned experiments like AMS-02 will have access to the full energy range of interest.
\end{abstract}
\maketitle

\newcommand{\B}{B^{(1)}}
\newcommand{\be}{\begin{equation}}
\newcommand{\ee}{\end{equation}}
\newcommand{\kref}[1]{(\ref{#1})}


\section{Introduction}

The observational evidence for the existence of dark matter has become overwhelming and covers an impressive range of distance scales, from galaxy rotation curves \cite{per} via the mass-to-light ratio of galaxy clusters \cite{bah} to large scale structure surveys \cite{lss}. Adding our knowledge from microwave background observations \cite{wmap}, supernova luminosity distances \cite{sn1a} and big bang nucleosynthesis (BBN), one can determine the dark matter density of the universe to a remarkable accuracy. Yet, the nature of the dark matter still remains a mystery. Weakly interacting massive particles (WIMPs), however, are particularly
attractive dark matter candidates, as gauge couplings and masses at the electroweak symmetry
breaking scale give a thermal relic abundance that naturally comes within the observed
range \cite{reviews}. 

A well-studied example of such a WIMP candidate is the neutralino, the lightest supersymmetric particle. We concentrate instead on an interesting alternative that has received much attention in recent years -- Kaluza-Klein (KK) dark matter that appears in models with large extra dimensions \cite{app,che,ser}. In contrast to the neutralino, which is a Majorana fermion, this dark matter candidate is a massive vector particle, and consequently there is a rich, new phenomenology to explore that may well apply also to other dark matter candidates that are non-Majorana in nature. Observational prospects have been explored in some detail \cite{obs_gen} and found to be promising, both for direct-detection experiments with very sensitive detectors \cite{direct}, as well as for indirect detection through a contribution to the cosmic ray flux in photons \cite{BBEGa,BBEGb}, positrons \cite{pos} and neutrinos \cite{hoob} of all energies up to the dark matter particle's mass. 

Here, we present the first analysis of the expected contribution to the antiproton flux. Compared to supersymmetry \cite{BEU,don,LMZa}, the a priori interesting feature of KK dark matter in this respect is a rather large branching ratio into quarks and gluons, and a mass at the TeV scale, which is higher than what is usually the case for neutralinos. Consequently, the focus of our attention will be on a possible distortion of the antiproton spectrum for high energies (above several GeV). Traditionally, one has instead often focused on low energies (below and around the peak at about 1 GeV) when looking for exotic physics in the antiproton spectrum, which has the disadvantage of being plagued by much greater uncertainties on the theoretical side.

This paper is organized as follows. We begin by introducing the model that gives rise to the KK dark matter candidate $\B$. In Section \ref{sec_ann} we then present the antiproton yield from $\B$ annihilations in the galactic halo. Since the expected flux depends to a large extent on the details of the dark matter distribution, different halo profiles will be considered in Section \ref{sec_halo}. This includes a detailed discussion of the possible existence of dark matter substructure (``clumps''), with updated bounds on the amount of clumpiness that is observationally allowed in these kinds of scenarios. The two-zone cylindrical diffusion model that is used to propagate the antiprotons through the galaxy is presented in Section \ref{sec_prop} and in Section \ref{sec_flux} the expected antiproton spectrum is shown and confronted with observational prospects. Finally, Section \ref{sec_conc} concludes.

\section{Universal extra dimensions}
\label{sec_ued}

The lightest KK particle (LKP) that appears in models of
universal extra dimensions (UED) \cite{app}, is the first viable particle dark
matter candidate to arise from extra-dimensional extensions of the
standard model (SM) of particle physics (for the first
proposal of TeV sized extra dimensions, see \cite{anton}). In UED models, all SM fields propagate in the higher-dimensional bulk. For the effective four-dimensional theory, this means that every particle is accompanied by a tower of increasingly more massive KK states. Momentum
conservation in the extra dimensions, which leads to
KK mode number conservation in the effective four
dimensional theory, is broken at the boundaries of the compact
space by an orbifold compactification that projects out unwanted degrees of freedom at the zero mode level. If the boundary terms introduced at the orbifold fixed points are identical, a remnant of KK mode number $n$ conservation is left
in the form of conserved KK parity, $(-1)^n$, and the LKP is stable \cite{che}. This is analogous to conserved
R-parity in supersymmetric models, which in that case ensures the stability of the
lightest supersymmetric particle.  An attractive feature of this dark matter candidate, as compared to the supersymmetric neutralino, is that the unknown parameter space is quite
small and will be scanned throughout by next generation's accelerator
experiments.

We will consider the simplest, five dimensional model with one UED
compactified on an $S^1/Z_2$ orbifold of radius $R$. At tree-level,
the $n$th KK mode mass is then given by
\be
  m^{(n)} = \sqrt{(n/R)^2 + m_\mathrm{EW}^2},
\ee
where $m_\mathrm{EW}$ is the zero mode mass of the corresponding SM particle. However, the identification
of the LKP is nontrivial because radiative corrections to the mass
spectrum of the first KK level are typically larger than the
corresponding electroweak mass shifts. A one-loop calculation
shows that the LKP is well approximated by the first KK
mode of the weak hypercharge gauge boson $\B$ \cite{che}. The $\B$ relic density was
determined in \cite{ser,kak}. Depending on the exact form of the mass
spectrum and the resulting coannihilation channels, the measurements from the
Wilkinson Microwave Anisotropy Probe (WMAP) \cite{wmap} of
$\Omega_\mathrm{CDM} h^2 = 0.12 \pm 0.03$ corresponds to $0.5\mathrm{~TeV}
\lesssim m_{\B} \lesssim 1\mathrm{~TeV}$ if all dark matter is made of LKPs. Here, $\Omega_\mathrm{CDM}$ is
the ratio of cold dark matter to critical density and $h$ is the Hubble constant in
units of $100\mathrm{~km} \mathrm{\,s}^{-1} \mathrm{\,Mpc}^{-1}$. Collider
measurements of electroweak observables give a current constraint of
$R^{-1} \gtrsim 0.3\mathrm{~TeV}$ \cite{app,aga}, whereas LHC should
probe compactification radii up to 1.5 TeV \cite{cheb}.

\section{Antiprotons from LKP annihilations}
\label{sec_ann}

The number of antiprotons $\bar p$ per unit time, energy and volume element that are produced by LKP annihilations at some position $\mathbf r$ in the galactic halo is given by the source function
\be
  \label{source}
  Q_{\bar p}(T,\mathbf r)=\frac{1}{2}\left<\sigma_\mathrm{ann} v\right>_{\B}\left(\frac{\rho_{\B}(\mathbf r)}{m_{\B}}\right)^2\sum_fB^f\frac{\mathrm{d}N^f}{\mathrm{d}T}\,,
\ee
where $T$ is the antiproton kinetic energy, $\rho_{\B}$ the $\B$ halo density and $m_{\B}$ its mass. The sum runs over all annihilation channels $f$, where $B^f$ and $\mathrm{d}N^f/\mathrm{d}T$ are the branching ratios and fragmentation functions into antiprotons, respectively. The factor $1/2$ appears due to the fact that $\B$ particles always annihilate in pairs; note that this factor has often not been taken into account when calculating antiproton (or other cosmic ray) yields. Finally, the $\B$ annihilation rate is given by \cite{ser} 
\be
 \left<\sigma_\mathrm{ann} v\right>_{\B}
\simeq 3\cdot10^{-26} (0.8\mathrm{~TeV}/m_{\B})^2 \mathrm{~cm}^3
\mathrm{\,s}^{-1}\,.
\ee

\begin{table}[t]
   \begin{tabular}{|l||c|c|c|} 
        \hline
    \textbf{Annihilation}  &  \multicolumn{3}{c|}{\textbf{Mass shift $\frac{M}{m_{\B}}$}}\\
           \textbf{channel}& $1.05$    &   $1.1$  & $1.2$    \\
        \hline \hline
       $\B\B\rightarrow t\bar t$  & ~10.77 [10.27, 10.88]~  & ~10.73 [10.20, 10.85]~ & ~10.66 [10.06, 10.79]~ \\
       $\B\B\rightarrow c\bar c$ & ~11.09 [11.13, 11.08]~ & ~11.08 [11.12, 11.07]~ & ~11.05 [11.10, 11.04]~   \\
       $\B\B\rightarrow b\bar b$  & ~0.73 [0.74, 0.73]~ & ~0.73 [0.74, 0.73]~ & ~0.73 [0.73, 0.73]~    \\
       $\B\B\rightarrow W^+W^-$  & ~0.94 [0.99, 0.93]~ & ~0.98 [1.03, 0.96]~ & ~1.08 [1.14, 1.06]~  \\
       $\B\B\rightarrow ZZ$  &  ~0.47 [0.51, 0.46]~ & ~0.49 [0.52, 0.48]~ & ~0.54 [0.58, 0.53]~  \\
       $\B\B\rightarrow HH$   & ~0.54 [0.62, 0.52]~ & ~0.59 [0.69, 0.57]~ & ~0.72 [0.84, 0.69]~  \\
        \hline
   \end{tabular}
  \caption{Branching ratios (in percent) for all annihilation channels at tree level that are relevant for antiproton production. The values shown above are calculated for LKP masses $m_{\B}=$ 0.8 [0.5, 1.0] TeV and several values for the mass shift between the LKP mass and the mass $M$ of the other first level KK modes.}
  \label{tab_bf}
\end{table}

A considerable source of astrophysical uncertainty lies in the unknown properties of the dark matter halo profile $\rho_\mathrm{CDM}$ (which in the following will be assumed to be dominated by LKPs, i.e.~$\rho_\mathrm{CDM}\simeq\rho_{\B}$); this will be discussed in detail in the next section. In this section, we will instead concentrate on the antiproton differential spectrum per annihilation event
\be
 \label{gt}
 g(T)\equiv\sum_fB^f\frac{\mathrm{d}N^f}{\mathrm{d}T}\,,
\ee
that can be computed to a much better accuracy. At tree-level, with all other
first level KK modes degenerate in mass, $\B$ pairs annihilate into
quark pairs (35\%), charged lepton pairs (59\%), neutrinos (4\%),
charged (1\%) and neutral (0.5\%) gauge bosons and Higgs bosons (0.5\%). 
Table \ref{tab_bf} shows the exact branching ratios for all final states that are relevant for antiproton production. Clearly, they are very insensitive to the parameters of the model.

The annihilation into gluon pairs is loop-suppressed, but since one can expect a rather large antiproton yield from gluon hadronization, it seems worthwhile to include this annihilation channel as well. Recently, the similar process $\B\B\rightarrow\gamma\gamma$ has been analysed in  detail \cite{BBEGb}. For a two-gluon final state, only the fermion box diagrams appearing in that analysis contribute, with of course no contribution from leptons. The only potential contribution from loops containing scalars comes from diagrams with an $s$-channel Higgs coupled to a quark loop (to which also the final gluons couple); these amplitudes, however, are suppressed by a factor $m_q/m_{\B}$ due to the Yukawa coupling and can thus be neglected \footnote{Doing the full analytical calculation it is found that, even for the top quark, this diagram contributes only 1.7\% to the cross-section of the full process $\B\B\rightarrow gg$.}.
One can therefore easily derive an \emph{exact} expression, up to $\mathcal{O}\left(m_q/m_{\B}\right)$, for the total cross section of $\B\B\rightarrow gg$: Just take Eq. (16) in \cite{BBEGb} and substitute the electric charge $Q$ with $1$ (this corresponds to setting $g^2_{eff}=11/2$); the color average is then performed by the simple recipe $\alpha_{em}^2\rightarrow (2/9)\,\alpha_s^2$ \cite{BS}. The result is a cross section for gluon pair production that is roughly 0.5\% of the total annihilation cross section at tree level, i.e.~about the same size as the (tree-level!) contributions from massive vector bosons.

\begin{figure}[t]
 \psfrag{0.01}[r][r][0.8]{$0.01$}
 \psfrag{0.1}[r][r][0.8]{$0.1$}
 \psfrag{1}[r][r][0.8]{$1$}
 \psfrag{10}[r][r][0.8]{$10$}
 \psfrag{100}[r][r][0.8]{$100$}
 \psfrag{1000}[r][r][0.8]{$1000$}
 \psfrag{x}[][][1]{$T$ [GeV]} 
 \psfrag{y}[][][1][90]{$T^3\, g(T)$ [$\mathrm{GeV}^2$]}
 \includegraphics[width=\textwidth]{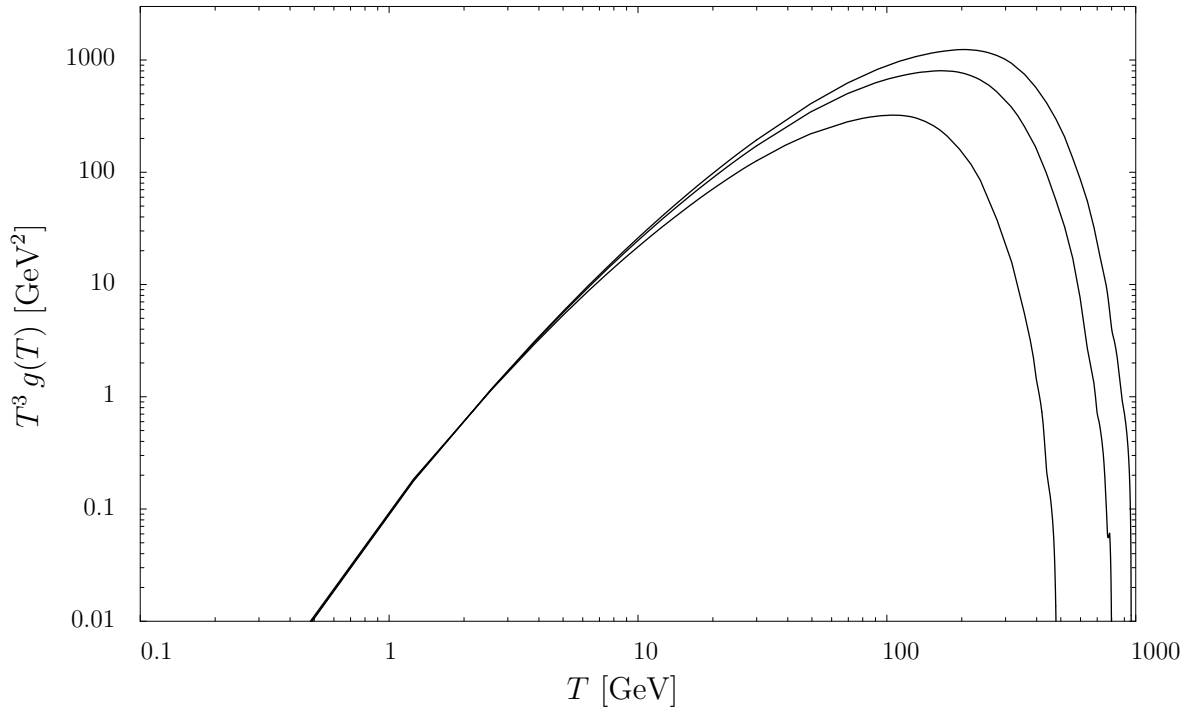}
 \caption{The differential antiproton spectrum per $\B$ pair annihilation event. From bottom to top, the plots correspond to an LKP mass of $m_{\B}=500$, 800 and 1000 GeV, respectively.}
 \label{gtfig}
\end{figure}

We use the tabulated fragmentation functions $\mathrm{d}N^f/\mathrm{d}T$ of the \textsc{DarkSUSY} package \cite{ds}, which are based on a large number of events ($10^6$ per annihilation channel and mass) as simulated with the Monte Carlo program \textsc{Pythia 6.115} \cite{pythia}. (While the parametrizations given in \cite{BEU} reproduce the spectrum to a very good accuracy for low and intermediate energies, they overpredict the highest-energy tail by a factor of up to 3). The fragmentation function for Higgs final states is not explicitly given; it can easily be obtained by letting the Higgs decay in flight and then boosting the fragmentation functions for those decay products that are relevant for antiproton production. The resulting total differential antiproton spectrum $g(T)$ is shown in Fig.~\ref{gtfig} for different $\B$ masses. Analyzing the individual terms in the sum (\ref{gt}) separately, one finds that top and charm quarks give as expected the largest contribution, except for the highest energies, where the $W^+W^-$ channel starts to dominate.

\section{Halo profiles}
\label{sec_halo}

As described in the last section, the ingredients for the antiproton source function $Q_{\bar p}$ that derive from the underlying particle physics can be computed in a straight-forward way to a satisfying degree of accuracy. Unfortunately, the same does not hold for the astrophysical part; the distribution of dark matter in our galaxy is to a large extent unknown and since $\rho_\mathrm{CDM}$ enters $Q_{\bar p}$ quadratically, this puts considerable limits on the possibility of making reliable predictions for the expected flux of cosmic rays.

\subsection{Smooth dark matter distributions}
\label{sec_smooth}

Direct observations can only poorly constrain the form of the halo profile in the Milky Way, so it is usually taken from N-body simulations of gravitational clustering. Since these simulations only reach resolutions of about $0.1$ kpc for galaxies with the size of the Milky Way, the innermost slope of the density profile has to be extrapolated and is thus bound to a considerable amount of theoretical uncertainty. Having this in mind, a generic parameterization of possible spherical halo profiles is given by
\be
  \label{prof}
  \rho_\mathrm{CDM}(\mathbf{r})=\rho_0\left(\frac{R_0}{r}\right)^\gamma\left[\frac{1+(R_0/a)^\alpha}{1+(r/a)^\alpha}\right]^\frac{\beta-\gamma}{\alpha}\,,
\ee
where the indices $(\alpha,\beta,\gamma)$ fix the halo model in question, $R_0=8.5$~kpc is the distance of the sun to the galactic center and $\rho_0\sim0.3$~GeV~cm$^{-3}$ the local halo density, which is known only up to a factor of almost 2 \cite{bel}. The scale radius $a$ determines where the transition for the radial dependence of $\rho_\mathrm{CDM}$ from large ($\rho_\mathrm{CDM}\propto r^{-\beta}$) to small ($\rho_\mathrm{CDM}\propto r^{-\gamma}$) galactocentric distances takes place. N-body simulations favour cuspy halo distributions like the NFW profile \cite{nfw} with $(\alpha,\beta,\gamma)=(1,3,1)$, or the even more cuspy Moore profile \cite{moore} with $(1.5,3,1.5)$. In fact, there is also analytic support for considering profiles with $\gamma\geq1$ \cite{han}.
Here, we will for comparison also take into account the rather conservative case of an isothermal sphere, $(2,2,0)$, that has a constant density core. There are indications that the scale radius $a$ is not an independent parameter but strongly correlated with the virial mass of the galaxy \cite{eke}. In the case of the Milky Way and the isothermal sphere (NFW, Moore) profile, one finds $a=4$ kpc (21.7 kpc, 34.5 kpc) \cite{lidia}, which we will use in the following as reference values. 

The most recent simulations actually suggest a universal halo profile of the form
\be
  \label{exphalo}
 \rho_\mathrm{CDM}(\mathbf{r})\approx\rho_0\exp\left[-\left(r/a\right)^\frac{1}{n}\right],
\ee
with $n\approx5$ \cite{Nbody}. However, taking into account the effects of baryons in a simple model of adiabatic contraction \cite{blum}, this profile is effectively transformed to a Moore profile \cite{mooretrafo}. Therefore, we restrict ourselves in the following to the isothermal sphere, NFW and Moore profiles. This should cover a plausible range of realistic halo models, serving to illustrate the uncertainties that derive from our lack of knowledge of the exact form of the galactic dark matter distribution.

\subsection{Dark matter substructures}
\label{sec_clumps}

In cosmologies with cold dark matter, structure forms hierarchically,
so it is reasonable to assume the existence of substructures
(``clumps'') within the smooth halo distributions discussed in the
last section. This is very interesting from the point of view of
cosmic rays as an indirect detection method of dark matter, since the
total expected flux is always obtained by averaging
$\rho_\mathrm{CDM}^2$ in one way or the other -- which generically
gives larger values for inhomogeneous dark matter distributions. This
effect has mainly been studied for substructures larger than $10^6
M_\odot$, roughly correponding to the resolution limit of the N-body
simulations mentioned earlier. Recently however, attention has turned
to much smaller scales when it was discovered that for every WIMP dark
matter candidate there is a natural cut-off $M_c$ in the power
spectrum, below which density perturbations are washed out by
collisional damping and subsequent free streaming \cite{stefan}. A
considerable part of the dark matter should therefore have collapsed
into substructures with a mass of about $M_c$, thereby forming the
first gravitationally bound objects in the cosmological
evolution. Based on very recent, high-resolution N-body simulations,
it was estimated that a relatively large fraction of these microhalos
can survive until today, giving a contribution of several percent to the total amount of dark matter in the solar neighborhood \cite{die}. These results are based on the assumption of neutralino dark matter, so eventually the whole analysis has to be redone for the case of KK dark matter \cite{BH}; while this can certainly change details like the value of $M_c$, the main results -- i.e.~the existence of a cutoff and a rather large survival probability of the first gravitationally bound objects -- should persist.

Here, we will follow the approach of \cite{oda} and assume that the microhalos trace the mass distribution except for regions very close to the galactic center, where strong tidal forces and/or interaction with stars are expected to have destroyed all substructure by now. A very simplified model to describe the clumpiness of any halo distribution then contains three parameters; the disruption radius $R_d$ within which all substructure is completely disrupted, the fraction $f$ of the dark matter outside this radius that comes in clumps, and the dimensionless parameter
\be
  \label{profiles}
  \delta\equiv\frac{1}{\rho_0}\frac{\int\mathrm{d}^3\mathbf{r}_\mathrm{cl}\left(\rho_\mathrm{cl}(\mathbf{r}_\mathrm{cl})\right)^2}{\int\mathrm{d}^3\mathbf{r}_\mathrm{cl}\,\rho_\mathrm{cl}(\mathbf{r}_\mathrm{cl})}\,,
\ee
which describes the effective contrast between an average dark matter clump and the local halo density. The numerical results indicate microhalo profiles $\rho_\mathrm{cl}\propto r^{-\gamma}$ with no further substructure and slopes $\gamma$ in the range from 1.5 to 2 down to the resolution limit of the simulation \cite{die}. Taking the average slope to be 1.7, one can in principle obtain density contrasts up to $\delta\sim10^6$ in the most extreme case, depending on how far one trusts the extrapolation. (Here, we have assumed the existence of a minimal radius inside which the density $\rho_\mathrm{cl}$ is constant and the inverse of the self-annihilation rate, $\left[\left<\sigma_\mathrm{ann} v\right>_{\B}\rho_\mathrm{cl}/m_{\B}\right]^{-1}$, equals the typical time-scale for the formation of the singularity, $\left(G\bar\rho_\mathrm{cl}\right)^{-\frac{1}{2}}$, where $\bar\rho_\mathrm{cl}$ is the average clump density at collapse \cite{ber}).

To calculate the collective contribution from clumps to the total annihilation flux, one just has to replace
\be
 \label{r2clump}
 \left(\rho_\mathrm{CDM}(\mathbf{r})\right)^2\rightarrow f\delta\rho_0\,\rho_\mathrm{CDM}(\mathbf{r})\Theta(r-R_d)
\ee
in the source function (\ref{source}). The only assumption that enters here is that the number density of clumps in the galaxy, tracing the halo profile, is essentially constant over regions with the size of an individual clump -- which should be a good approximation for almost any scenario. The simple recipe (\ref{r2clump}) therefore applies not only to antiprotons, but to \emph{any} final state of dark matter annihilations.

With a clumpy halo distribution, any contribution of dark matter annihilation products to cosmic rays will be enhanced. Therefore one must be cautious not to violate observational bounds on the flux; this puts severe constraints on the allowed parameter space of the clumpy halo model described above. For particles with a long free pathlength, however, the by far greatest flux is still expected from the very dense region around the galactic center where there is no enhancement due to the presence of clumps. To illustrate this, let us consider the case of photons, where the flux observed with a detector of angular acceptance $\Delta\Omega$ is proportional to a line of sight integral towards the galactic center, $J\equiv\int_{\Delta\Omega}\!\mathrm{d}\Omega\int_{l.o.s.}\!\mathrm{d}l\,\rho_\mathrm{CDM}(\mathbf{r})^2$. The contribution $J_\mathrm{cl}$ from the clumps alone is therefore given by
\be
  J_\mathrm{cl}/\bar J= f \delta\, \rho_0\frac{\int\!\mathrm{d}\Omega\int\!\mathrm{d}l\,\bar \rho_\mathrm{CDM}(\mathbf{r})\Theta(r-R_d)}{\int\!\mathrm{d}\Omega\int\!\mathrm{d}l\,\bar \rho_\mathrm{CDM}(\mathbf{r})^2}\equiv f \delta\, b_\mathrm{cl}^\gamma\,,
\ee
where $\bar J\equiv J\big|_{\rho_\mathrm{CDM}\rightarrow\bar \rho_\mathrm{CDM}}$ and $\bar \rho_\mathrm{CDM}$ is the smooth halo profile parametrized by (\ref{profiles}). An NFW profile with $R_d=1$ kpc, for example, gives $b_\mathrm{cl}^\gamma\approx 6\cdot10^{-4}$, which means that the continuous gamma ray signal from KK dark matter annihilations \cite{BBEGa} in that case does not overpredict the data taken by HESS \cite{HESS} (or EGRET \cite{EGRET}) for clump density contrasts as high as $\delta\lesssim10^5/f$ ($\delta\lesssim10^6/f$). 

\section{Propagation through the galaxy}
\label{sec_prop}

For the propagation of antiprotons through the galaxy we use the semiempirical two-zone cylindrical diffusion model of ref.~\cite{BEU} to determine the expected number density $N_{\bar p}(T,\mathbf{r})$ of antiprotons from LKP annihilations, given the source function (\ref{source}). In this approach, the galaxy is split into two parts and described as a thin disk of radius $R_h=20$~kpc and height $2h_g=0.2$~kpc that contains the interstellar gas (with number density $n_g=1$~cm$^{-3}$), and a halo of the same radius and height $2 h_h$ (which is not equally well known). To simplify things, we assume the same diffusion coefficient
\be
  D(\mathcal{R})=D_0\left(1+\frac{\mathcal{R}}{\mathcal{R}_0}\right)^\delta
\ee
for both parts, with a dependence on the particle rigidity $\mathcal{R}$ (i.e. the momentum of the particle per unit charge) as indicated. Antiproton losses due to inelastic scattering with the interstellar gas, primarily hydrogen, are taken into account by a corresponding loss term in the diffusion equation. Given the cylindrical symmetry, and the boundary condition that antiprotons can escape freely at the border of the propagation region, it is then possible to find an analytical solution to the problem; its explicit form, expressed as an expansion in Bessel functions, is given in \cite{BEU}.

One may refine this simple model by allowing for the presence of a
galactic wind, blowing particles away from the disk, or by taking into
account the effect of the (time-varying) solar activity on charged
particles near earth. Another possible  effect is reacceleration by
stochastic magnetic fields during propagation or shock waves from
supernova remnants. In the very high-energy regime that we are
interested in, however, these effects are of minor importance (see \cite{BEU,
don} and references therein) and we do not consider them here.

The allowed range of propagation parameters can be constrained by analyzing the flux of stable nuclei, mainly by fitting the observed boron to carbon ratio B/C. This has been done in detail for a very similar propagation model \cite{mau} and in the following we will take a choice of parameters that results in a median flux of antiprotons \cite{don01}: $h_h=4$~kpc, $\delta=0.7$ and $D_0=3.5\cdot10^{27}$~cm$^2\,$s$^{-1}$. The cutoff scale is set to $\mathcal{R}_0=1$~GV; again, its exact value does not influence the antiproton flux at high energies.

\section{Expected antiproton fluxes}
\label{sec_flux}

Having solved the diffusion equation for the number density $N_{\bar p}(T,\mathbf{r})$, the expected flux of antiprotons from LKP annihilations is readily obtained as
\be
  \Phi_{\bar p}(T)=\frac{v_{\bar p}}{4\pi}N_{\bar p}(T,R_0)\,.
\ee
In addition to this primary source of antiprotons, there is a background contribution to the flux that is dominated by antiprotons from inelastic cosmic ray collisions with the interstellar medium, mostly protons colliding with hydrogen and helium. Since the spectra of both cosmic ray protons and helium nuclei are quite well determined, one can again use the B/C ratio to determine the propagation parameters describing the diffusion model and obtain a prediction for the background flux that fits the available antiproton data remarkably well \cite{maub}.

\begin{figure}[t]
\psfrag{m5}[][][0.8]{$10^{-5}$}
\psfrag{m3}[][][0.8]{$10^{-3}$}
\psfrag{m1}[][][0.8]{$10^{-1}$}
\psfrag{p1}[][][0.8]{$10^{1}$}
\psfrag{p3}[][][0.8]{$10^{3}$}
\psfrag{1}[][][0.8]{$1$}
\psfrag{10}[][][0.8]{$10$}
\psfrag{100}[][][0.8]{$100$}
\psfrag{1000}[][][0.8]{$1000$}
\psfrag{l1}[l][l][0.9]{BESS 1995-1997}
\psfrag{l2}[l][l][0.9]{BESS 1998}
\psfrag{l3}[l][l][0.9]{CAPRICE 1998}
\psfrag{x}[][][1]{$T$ [GeV]}
\psfrag{y}[][][1][90]{$T^3 \Phi_{\bar p}$ [$\mathrm{m}^{-2}\,\mathrm{s}^{-1}\,\mathrm{sr}^{-1}\,\mathrm{GeV}^2$]}
 \includegraphics[width=\textwidth]{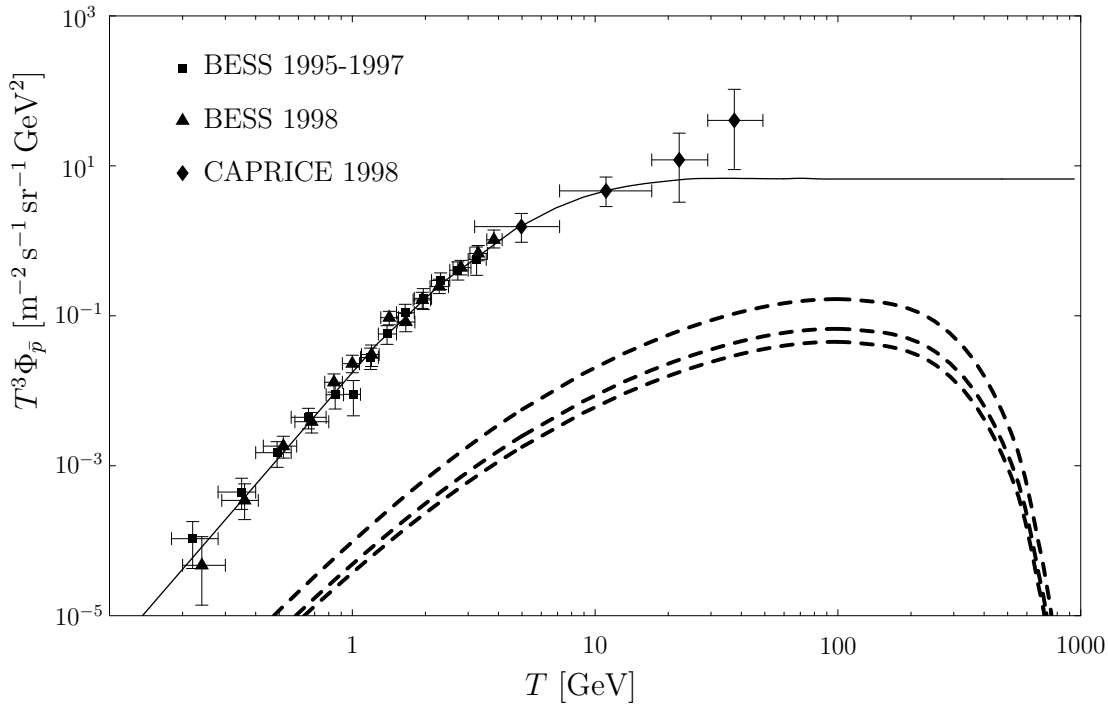}
 \caption{The antiproton background flux as determined in \cite{maub}, extrapolated for the energy range from 100 GeV to 1 TeV, is shown together with the data taken by the balloon-borne experiments BESS \cite{bess} and CAPRICE \cite{caprice}. The dashed lines show the contribution from LKP annihilations, with an LKP mass of $m_{\B}=800$~GeV and, from bottom to top, an isothermal, NFW and Moore halo profile, respectively.}
 \label{flux_profiles}
\end{figure}
\begin{figure}[t]
\psfrag{m5}[][][0.8]{$10^{-5}$}
\psfrag{m3}[][][0.8]{$10^{-3}$}
\psfrag{m1}[][][0.8]{$10^{-1}$}
\psfrag{p1}[][][0.8]{$10^{1}$}
\psfrag{p3}[][][0.8]{$10^{3}$}
\psfrag{1}[][][0.8]{$1$}
\psfrag{10}[][][0.8]{$10$}
\psfrag{100}[][][0.8]{$100$}
\psfrag{1000}[][][0.8]{$1000$}
\psfrag{x}[][][1]{$T$ [GeV]}
\psfrag{y}[][][1][90]{$T^3 \Phi_{\bar p}$ [$\mathrm{m}^{-2}\,\mathrm{s}^{-1}\,\mathrm{sr}^{-1}\,\mathrm{GeV}^2$]}
 \includegraphics[width=\textwidth]{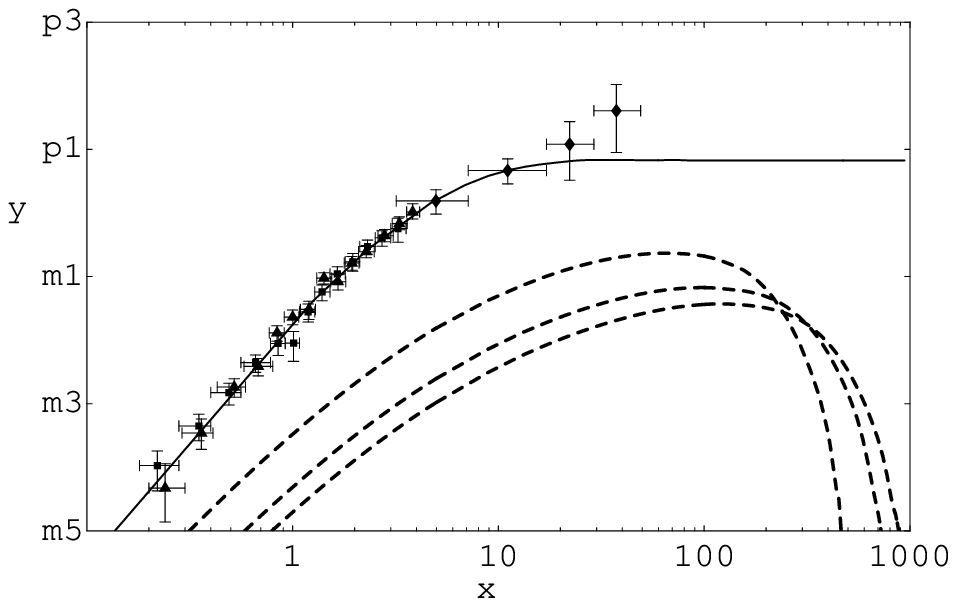}
 \caption{The same as Fig.~\ref{flux_profiles}, now for an NFW profile and, from left to right, LKP masses $m_{\B}=500$, 800 and 1000 GeV.}
 \label{flux_masses}
\end{figure}

This background flux is shown in Fig.~\ref{flux_profiles}, together with the primary component from LKP annihilations. As one can see, the dependence on the halo profile is much smaller than for cosmic ray photons, where the induced uncertainties in the flux span several orders of magnitude \cite{BUB,lidia}; this can easily be understood by noting that the halo distributions in question mainly differ very near the galactic center -- whereas most antiprotons simply do not reach that far through the diffusive halo 
\footnote{
Note, however, that we have adopted here a rather conservative version of the Moore profile and assumed a constant density inside a core radius of 0.1 kpc, which is roughly the resolution of the underlying N-body simulations. Extrapolating the $r^{-1.5}$ behaviour down to $10^{-8}$ kpc, where higher densities are prevented by LKP self-annihilation \cite{ber}, gives an enhancement of 5.3 (for low energies) up to 10.5 (for high energies) as compared to the NFW case
.}
. A much greater astrophysical uncertainty is connected to the fact that the B/C analysis actually allows a rather large range of propagation parameters; even though this results in nearly degenerate background predictions,
 the flux from primary sources can vary almost an order of magnitude at higher energies (for energies below about 1~GeV the induced uncertainy spans two orders of magnitude) \cite{don01,don}. A thorough analysis of the uncertainties related to the choice of propagation model, however, is beyond the scope of this paper and therefore we have adopted a choice of parameters that results in a median flux of primary antiprotons, as specified in Section \ref{sec_prop}. Finally, the uncertainty in the local halo density $\rho_0$ induces another factor of about 4 for the primary flux. For completeness, Fig. \ref{flux_masses} shows the dependence of the primary flux on the LKP mass.

\begin{figure}[t]
\psfrag{0.1}[r][r][0.8]{$0.1$}
\psfrag{1}[r][r][0.8]{$1$}
\psfrag{10}[r][r][0.8]{$10$}
\psfrag{100}[r][r][0.8]{$100$}
\psfrag{1000}[r][r][0.8]{$1000$}
\psfrag{l1}[l][l][0.9]{AMS-02}
\psfrag{l2}[l][l][0.9]{PAMELA}
\psfrag{l3}[l][l][0.8]{(projected 3 year data)}
\psfrag{x}[][][1]{$T$ [GeV]}
\psfrag{y}[][][1][90]{$T^3 \Phi_{\bar p}$ [$\mathrm{m}^{-2}\,\mathrm{s}^{-1}\,\mathrm{sr}^{-1}\,\mathrm{GeV}^2$]}
 \includegraphics[width=\textwidth]{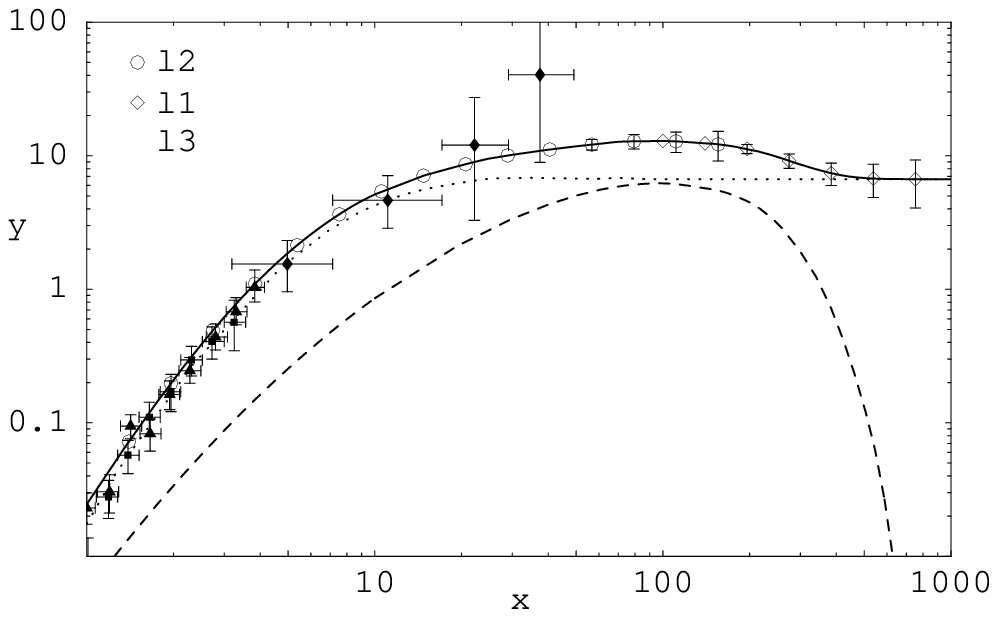}
 \caption{The solid line shows the expected antiproton spectrum for the case of a clumpy NFW profile with $f\delta=200$ and $m_{\B}=800$~GeV; the dotted and dashed lines give, respectively, the background flux and the contribution from LKP annihilations alone. The data points are the same as those of Fig.~\ref{flux_profiles}; in addition, the detectional prospects of PAMELA \cite{pamela} and AMS-02 \cite{ams} are indicated by displaying their projected data after three years of operation (only statistical errors are included; error bars smaller than the symbol size are not shown). For AMS-02, only energies above 100 GeV are considered.}
 \label{flux_clump800}
\end{figure}

Comparing the expected background and primary fluxes, prospects for the indirect detection of KK dark matter through its annihilation into antiprotons clearly do not look very promising. The situation changes drastically, however, if one allows for clumpy halo distributions. Making use of (\ref{r2clump}), the contribution from the clumps alone may be expressed as
\be
  \Phi_{\bar p}^{cl}= f\delta\, b_\mathrm{cl}^{\bar p}~ \overline{\Phi}_{\bar p},
\ee
where $\overline{\Phi}_{\bar p}$ is the antiproton flux for the smooth halo distribution. The formfactor $b_\mathrm{cl}^{\bar p}$ decreases with energy and varies from 0.99 (0.61, 0.35) at $T=0.1$ GeV to 0.89 (0.47, 0.20) at $T=30$ GeV for an isothermal (NFW, Moore) profile; above this, it stays approximately constant. As an example, Fig.~\ref{flux_clump800} shows the resulting flux for an NFW profile with $f\delta=200$ and an LKP mass $m_{\B}=800$~GeV (a Moore profile with $f\delta=$190  would give an almost identical spectrum). For such boost factors, an interesting spectral feature thus appears at energies right above the currently available data, that can clearly be distinguished from the background. As also indicated in the figure, the soon to be launched satellite experiment PAMELA \cite{pamela}, with its energy range up to 190 GeV, as well as the AMS-02 experiment \cite{ams}, planned to be installed on the international space station by 2007,  will have enough resolution to see -- or to rule out -- such a distortion in the spectrum. Already now, one may use the existing measurements in the low-energy range to put an upper bound on the allowed amount of halo clumpiness for KK dark matter that is much tighter than the bound obtained from gamma rays. In the case of an LKP mass of 800 GeV, one has to roughly satisfy $f\delta\lesssim10^3$ in order not to overpredict the antiproton flux that is measured at energies below 1~GeV -- though a careful analysis of this bound should of course take into account the much higher uncertainties connected to the expected primary flux at low energies.

\begin{figure}[t]
\psfrag{0.1}[r][r][0.8]{$0.1$}
\psfrag{1}[r][r][0.8]{$1$}
\psfrag{10}[r][r][0.8]{$10$}
\psfrag{100}[r][r][0.8]{$100$}
\psfrag{1000}[r][r][0.8]{$1000$}
\psfrag{x}[][][1]{$T$ [GeV]}
\psfrag{y}[][][1][90]{$T^3 \Phi_{\bar p}$ [$\mathrm{m}^{-2}\,\mathrm{s}^{-1}\,\mathrm{sr}^{-1}\,\mathrm{GeV}^2$]}
 \includegraphics[width=\textwidth]{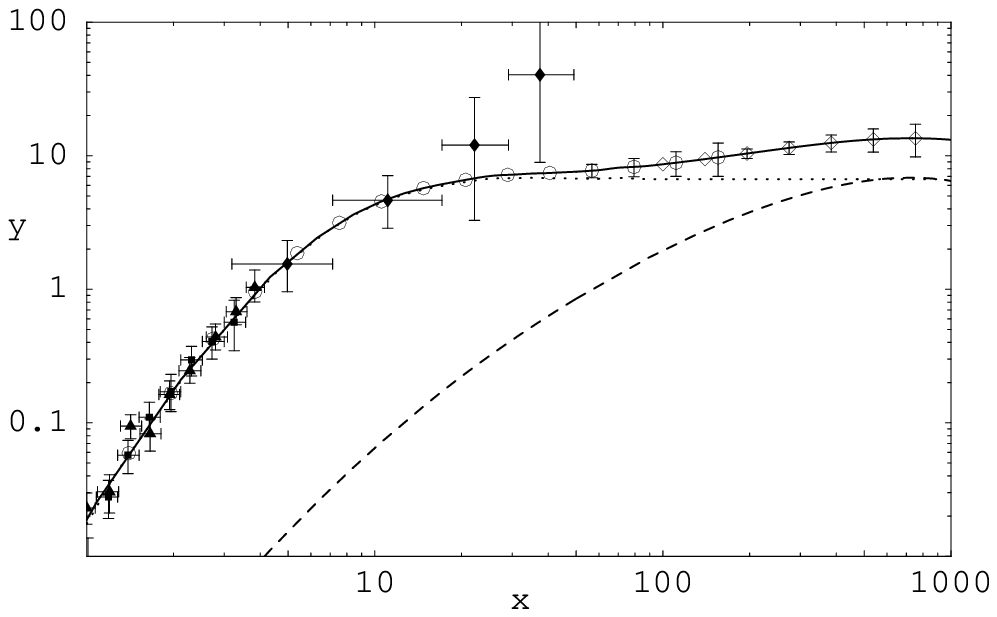}
 \caption{The same as in Fig.~\ref{flux_clump800}, but now for $f\delta=$1000 and a hypothetical  dark matter candidate with mass 7 TeV that was recently proposed to explain the observed gamma ray signal from the galactic center \cite{BBEGa}.}
 \label{flux_clump7000}
\end{figure}

As a further example of how the distortion of the antiproton spectrum for high energies might look like, Fig.~\ref{flux_clump7000} shows the case of a hypothetical dark matter candidate with mass 7 TeV and similar couplings like the LKP, that was recently proposed as a possible explanation for the TeV gamma ray signal from the galactic center \cite{BBEGa}. Due to the high mass, however, one needs even larger boost factors to see an effect at all.

\section{Conclusions}
\label{sec_conc}

As a massive vector particle, the LKP offers an interesting alternative to the most often studied dark matter candidate, the supersymmetric neutralino, which is a Majorana fermion. In this article, we have studied the contribution to the cosmic ray spectrum from LKP annihilations in our Galaxy, focusing on high-energetic antiprotons.

We found that the expected antiproton signal is dominated by the background flux for standard assumptions about (smooth) halo profiles. However, if one allows for clumpy halo distributions, which is supported by recent theoretical as well as numerical analyses \cite{stefan,die}, one can obtain a signal at high energies that clearly can be distinguished from the background flux already by the shortly upcoming experiment PAMELA \cite{pamela}; the AMS-02 experiment \cite{ams} will eventually be able to measure the full energy range of interest. One might worry that the required boost factors to obtain such a signal are uncomfortably high -- but it is important to note that existing numerical simulations actually do allow for very high values of the density contrast inside the clumps (keeping in mind, though, that they do not yet reach the resolutions that would be needed to finally settle this question; eventually, one also has to redo the analysis of \cite{stefan,die} for the specific case of KK dark matter \cite{BH}). As we have pointed out, even very high density contrasts need not violate bounds from gamma ray measurements, which might be the greatest concern; in fact, much more stringent constraints come from the available data on the (low energy) antiproton flux itself. We have also checked that the expected signal from positrons \cite{pos}, for the masses and boost factors considered here, is not in conflict with the available data.
Finally, the boost factors due to clumps that are required in the supersymmetric case are often even higher \cite{susyclumps}.

A full analysis of the astrophysical uncertainties, most notably the
impact of different propagation models on the results presented here,
is beyond the scope of this article and left open for future
studies. Another aspect well worth studying is the flux from  nearby
individual clumps; this would add to the average effect of small
clumps that we have considered in this article and thus further
enhance the signal. Furthermore, it should be of great interest to
carefully compare the predicted spectra for the LKP and the
neutralino, respectively, and see whether PAMELA or AMS-02 would be
able to distinguish these two cases. In this context it is an
important observation that the boost factors required for an
LKP-induced distortion of the antiproton spectrum \emph{automatically}
result in a distinguished peak signal in the \emph{positron} spectrum even
for the relatively high LKP masses considered here \cite{pos} -- while
such a signal would be absent in the supersymmetric case due to the
helicity-suppressed annihilation of  neutralinos into lepton
pairs. As an additional remark, the simultaneous appearance of the
positron peak in the case of a bump-like distortion of the antiproton spectrum would also
provide a rather firm evidence for the dark matter nature of the
latter and disfavour any attempt to merely attribute it to a lack of
understanding of the antiproton background flux.
 In fact, the positron excess reported by the HEAT experiment
\cite{heat} -- if confirmed by PAMELA and AMS-02 -- might already be
an indication for such a feature in the positron spectrum.
The high-energy window from about 10 GeV to several 100 GeV in the antiproton -- as well as positron -- spectrum will in any case remain a promising place to look for signatures from new physics.
\\

\ack

It is a pleasure to thank Lars Bergstr\"om, Martin Eriksson, Malcolm Fairbairn and Michael Gustafsson for useful discussions and comments on the manuscript, Joakim Edsj\"o for support with \textsc{DarkSUSY}, Fiorenza Donato for providing me with the data for the background flux, as well as Mirko Boezio, Silvio Orsi and Fernando Barao for detailled information about the PAMELA and AMS-02 experiments.
\vspace{2ex}

{\bf Final note:} During the last stages of this work, I became aware of a similar project on antiprotons from KK dark matter that was independently carried out by A. Barrau \emph{et al.} and appeared on the arXiv soon after the present article \cite{pierre}.


\end{document}